\documentclass[aps,prb,twocolumn]{revtex4-1}
\usepackage{amsmath,amssymb,bm,epsfig,graphicx,longtable}
\begin{document}

\title{Orbital-free approximations to the kinetic-energy density in
exchange-correlation MGGA functionals: tests on solids}
\author{Fabien Tran}
\author{P\'{e}ter Kov\'{a}cs}
\author{Leila Kalantari}
\author{Georg K. H. Madsen}
\author{Peter Blaha}
\affiliation{Institute of Materials Chemistry, Vienna University of Technology,
Getreidemarkt 9/165-TC, A-1060 Vienna, Austria}

\begin{abstract}

A recent study of Mejia-Rodriguez and Trickey [Phys. Rev. A \textbf{96},
052512 (2017)] showed that the deorbitalization procedure
(replacing the exact Kohn-Sham kinetic-energy density by an approximate orbital-free
expression) applied to exchange-correlation
functionals of the meta-generalized gradient approximation (MGGA) can lead
to important changes in the results for molecular properties.
For the present work, the deorbitalization of MGGA functionals is
further investigated by considering various properties of solids.
It is shown that depending on the MGGA,
common orbital-free approximations to the kinetic-energy density can be
sufficiently accurate for the lattice constant, bulk modulus, and cohesive energy.
For the band gap, calculated with the modified Becke-Johnson MGGA potential,
the deorbitalization has a larger impact on the results.

\end{abstract}

\maketitle

\section{\label{introduction}Introduction}

Kohn-Sham density functional theory\cite{HohenbergPR64,KohnPR65} (KS-DFT) is
a computationally efficient quantum method, which allows the treatment of
molecules, surfaces, and solids containing up to several thousands of atoms.
KS-DFT is particularly fast when the exchange and correlation (xc) effects are
treated at the semilocal level of approximation. The drawback is, however, that
there can be some degree of uncertainty in the results with semilocal
methods.\cite{CohenCR12,BeckeJCP14}
The most simple semilocal functional $E_{\text{xc}}$
is the local density approximation (LDA),\cite{KohnPR65,VoskoCJP80,PerdewPRB92a}
which is purely a functional of the electron density
$\rho=\sum_{i=1}^{N}\left\vert\psi_{i}\right\vert^{2}$.
Higher accuracy can be obtained by using functionals of the
generalized gradient approximation (GGA)\cite{BeckePRA88,LeePRB88,PerdewPRL96,PerdewPRL08}
which depend additionally on the first derivative of $\rho$ ($\nabla\rho$).
Nowadays, the most advanced and accurate semilocal functionals
are the so-called meta-GGA (MGGA),\cite{DellaSalaIJQC16} which, in addition
of $\rho$ and $\nabla\rho$, depend also on the
positive-definite KS kinetic-energy density (KED)
\begin{equation}
t^{\text{KS}}(\mathbf{r})=\frac{1}{2}\sum_{i=1}^{N}\nabla\psi_{i}^{*}(\mathbf{r})\cdot\nabla\psi_{i}(\mathbf{r})
\label{tKS}
\end{equation}
and/or the
second derivative of $\rho$ ($\nabla^{2}\rho$):
\begin{equation}
E_{\text{xc}}^{\text{MGGA}} =
\int\varepsilon_{\text{xc}}\left(\rho(\mathbf{r}),\nabla\rho(\mathbf{r}),
\nabla^{2}\rho(\mathbf{r}),t^{\text{KS}}(\mathbf{r})\right)d^{3}r.
\label{ExcMGGA}
\end{equation}
Considering how constructing a functional using more ingredients brings more
flexibility to it, MGGA functionals should
be universally more accurate than LDA and GGA functionals.
As with GGA functionals, a plethora of MGGA functionals have been proposed
(see Ref.~\onlinecite{DellaSalaIJQC16} for an exhaustive list) and
among the recent ones, SCAN\cite{SunPRL15} and TM\cite{TaoPRL16} for instance, have
shown to be accurate for many types of systems and properties.
\cite{TranJCP16,PengPRX16,MoJCP16,MoPRB17,HinumaPRB17}

As discussed in detail in Ref.~\onlinecite{DellaSalaIJQC16}, most
MGGA functionals depend only on the KED $t^{\text{KS}}$, while only very
few use also (or only)
$\nabla^{2}\rho$. One of the main reasons for not using $\nabla^{2}\rho$ in
$E_{\text{xc}}$ are the difficulties encountered when calculating
the potential (i.e., the functional derivative of $E_{\text{xc}}$)
for self-consistent calculations.
Indeed, the presence of $\nabla^{2}\rho$ in $E_{\text{xc}}$ means that the
potential contains a term, $\nabla^{2}\left(\partial\varepsilon_{\text{xc}}/
\partial\left(\nabla^{2}\rho\right)\right)$, that involves the third and fourth
derivatives of $\rho$ (see Ref.~\onlinecite{JemmerPRA95}) which
may lead to numerical problems like a greater sensitivity to the
integration grid.
\cite{JemmerPRA95,NeumannCPL97,CancioIJQC12,MejiaRodriguezPRA17}
(To our knowledge, only Ref.~\onlinecite{LaricchiaJCTC14}
reports an implementation of $\nabla^{2}\rho$-MGGA
with integration by part of the relevant Hamiltonian matrix elements\cite{NeumannMP96}
to avoid the third and fourth derivatives of $\rho$.)
As a comparison, a GGA potential involves only the first and second derivatives of
$\rho$ (or only the first if integration by part in the Hamiltonain
matrix\cite{PopleCPL92} is done),
and a $t^{\text{KS}}$-dependency in a MGGA functional leads to an additional
(non-multiplicative) term in the potential,
$-\left(1/2\right)\nabla\cdot\left(\left(
\partial\varepsilon_{\text{xc}}/\partial t^{\text{KS}}\right)\nabla\psi_{i}\right)$,
that involves the derivatives of $\psi_{i}$ up to the second
order (or only the first if integration by part in the Hamiltonian matrix\cite{NeumannMP96} is done).
Therefore, MGGA calculations have been done using mostly $t^{\text{KS}}$-MGGAs and
are becoming increasingly popular
(see Refs.~\onlinecite{FerrighiJCP11,SunPRB11b,WomackJCP16,YaoJCP17} for recent works
reporting self-consistent implementations for periodic solids).
Furthermore, from the theoretical point of view a benefit of using
$t^{\text{KS}}$ is that regions of space dominated by a single orbital
can be detected (see, e.g., Ref.~\onlinecite{BeckeJCP90}).

On the other hand, $\nabla^{2}\rho$-MGGAs have the advantage to be explicit
functionals of $\rho$ such that the functional derivative leads to a true
KS (i.e., multiplicative) potential, which is not the case with $t^{\text{KS}}$-MGGAs.
Also, except for the problems with the high derivatives of $\rho$ mentioned above,
a new self-consistent implementation of MGGAs should be easier for
$\nabla^{2}\rho$-MGGAs.
Thus, from the fundamental and practical point of views,
$\nabla^{2}\rho$-MGGAs are still of interest and worth to be further considered
as done in recent works.
\cite{CancioIJQC12,MejiaRodriguezPRA17,BienvenuJCTC18}

In particular, Mejia-Rodriguez and Trickey\cite{MejiaRodriguezPRA17}
investigated the effect of replacing the exact orbital-dependent $t^{\text{KS}}$
in existing $t^{\text{KS}}$-MGGA functionals by
some orbital-free (OF) approximations $t^{\text{OF}}$.
They called this procedure
\textit{deorbitalization}, meaning that a $t^{\text{KS}}$-MGGA is transformed into
an explicit density functional $\nabla^{2}\rho$-MGGA.
The properties that they considered are the heat of formation, bond lengths,
and vibration frequencies of molecules. This study showed
that the replacement $t^{\text{KS}}\rightarrow t^{\text{OF}}$
can have some impact on the results depending on the
xc-MGGA or the OF KED.
For instance, the average error for the heat of formation
is in some cases only slightly modified, while in some other cases
it is increased by one order of magnitude.
Also, it seems that none of the
OF KED they considered, including the two new ones proposed by
Mejia-Rodriguez and Trickey, leads to reasonably small changes in all cases.
For the present study, we pursue the investigations on the
deorbitalization procedure by considering properties of solids.
Several $t^{\text{KS}}$-MGGA energy functionals will be deorbitalized
and tested on the lattice constant, bulk modulus, and cohesive energy,
while the deorbitalization of the modified Becke-Johnson
potential\cite{TranPRL09} will be considered for the electronic structure.

The structure of the paper is the following. Section~\ref{theory} provides a
brief description of the theory and the computational details.
In Sec.~\ref{results}, the results obtained with the
deorbitalized MGGAs are presented and discussed, while
Sec.~\ref{discussion} provides some analysis and
Sec.~\ref{summary} gives the summary of this work.

\section{\label{theory}Theory and computational details}

\subsection{\label{kinetic}Orbital-free kinetic energy densities}

In the KS-DFT method,\cite{KohnPR65} the noninteracting kinetic energy component of
the total energy is given by
$T_{\text{s}}^{\text{KS}}=\int t^{\text{KS}}d^{3}r$, where $t^{\text{KS}}$ is
given by Eq.~(\ref{tKS}). Note that another common expression for the integrand in
$T_{\text{s}}^{\text{KS}}$ is 
$t^{\text{KS}'}=-\left(1/2\right)\sum_{i=1}^{N}
\psi_{i}^{*}\nabla^{2}\psi_{i}$ which is related to $t^{\text{KS}}$ by
$t^{\text{KS}'}=t^{\text{KS}}-\left(1/4\right)\nabla^{2}\rho$ and
leads to the same value of $T_{\text{s}}^{\text{KS}}$ since the integral
of $\nabla^{2}\rho$ is zero. For the development of fully OF DFT
methods\cite{Ligneres,Wesolowski13,KarasievAQC15} or in the framework
of embedding schemes,\cite{JacobWCMS14,WesolowskiCR15,SmigaJCP17,JiangJCP18}
expressions for $T_{\text{s}}$ which are
explicit functionals of $\rho$ have been proposed,
and as for xc-functionals, the majority of them are of semilocal type.
The most simple is the LDA of Thomas and Fermi\cite{ThomasPCPS27,FermiRANL27} (TF)
which is the exact expression for the homogeneous electron gas and reads
\begin{equation}
T_{s}^{\text{TF}} = C_{\text{TF}}\int\rho^{5/3}(\mathbf{r})d^{3}r,
\label{TsTF}
\end{equation}
where $C_{\text{TF}}=\left(3/10\right)\left(3\pi^{2}\right)^{2/3}$.
With respect to the exact values ($T_{\text{s}}^{\text{KS}}$),
the TF functional leads to underestimations for atoms\cite{Parr}
and molecules\cite{ThakkarPRA92,IyengarPRA01,TranCPL02,SeinoJCP18}
of about 10\%. Since the kinetic
energy is a major component of the total energy
$E_{\text{tot}}$ (from the virial theorem $T_{\text{s}}\approx-E_{\text{tot}}$)
such errors are extremely large.
Much better values for $T_{\text{s}}$ can be obtained with gradient-corrected
type (GGA) functionals
(errors below 0.5\% for the best ones
\cite{ThakkarPRA92,IyengarPRA01,TranIJQC02,TranCPL02,GarciaAldeaJCP07,SeinoJCP18}):
\begin{equation}
T_{s}^{\text{GGA}} = C_{\text{TF}}\int\rho^{5/3}(\mathbf{r})
F_{s}(s)d^{3}r,
\label{TsGGA}
\end{equation}
where
$s=\left\vert\nabla\rho\right\vert/\left(2\left(3\pi^{2}\right)^{1/3}\rho^{4/3}\right)$
is the reduced density gradient and $F_{s}$ is the kinetic enhancement factor for
which many forms have been proposed in the literature
(see Refs.~\onlinecite{KarasievJCAMD06,GarciaAldeaJCP07,Tran13,SeinoJCP18} for compilations)
like, for instance, those that were obtained using the \textit{conjointness conjecture}
between the exchange and kinetic energy functionals.\cite{MarchIJQC91,LeePRA91,TranIJQC02}
While GGAs can lead to rather accurate (albeit far from enough for
an useful OF DFT method) values of $T_{\text{s}}^{\text{GGA}}$,
the GGA KEDs defined as the integrand of
Eq.~(\ref{TsGGA}) show absolutely no resemblance to
Eq.~(\ref{tKS}).\cite{AlonsoCPL78,YangPRA86,FinzelTCA15,CancioJCP16} This can be understood
by considering the density-gradient expansion approximation (GEA)
of Eq.~(\ref{tKS}) which, at the second order,
is given by\cite{KirzhnitsJETP57,BrackPLB76} (L in GEA2L indicates the presence
of $\nabla^{2}\rho$)
\begin{equation}
t^{\text{GEA2L}}(\mathbf{r}) = t^{\text{TF}}(\mathbf{r}) +
\frac{1}{9}t^{\text{W}}(\mathbf{r}) + \frac{1}{6}\nabla^{2}\rho(\mathbf{r}),
\label{tGEA2L}
\end{equation}
where $t^{\text{TF}}=C_{\text{TF}}\rho^{5/3}$ [the integrand of Eq.~(\ref{TsTF})]
and $t^{\text{W}}=\left\vert\nabla\rho\right\vert^{2}/\left(8\rho\right)$ is the
von Weizs\"{a}cker\cite{vonWeizsackerZP35} KED. It is only by
considering $\nabla^{2}\rho$ in an OF KED $t^{\text{OF}}$ that
the shape of $t^{\text{OF}}$ can be made reasonably close to $t^{\text{KS}}$
(see Refs.~\onlinecite{AlonsoCPL78,YangPRA86,PerdewPRB07,SalazarIJQC16})
and despite some attempts,\cite{FinzelTCA15} it is most likely hopeless
to construct a GGA KED that looks similar to $t^{\text{KS}}$.

Thus, one has to consider $\nabla^{2}\rho$-dependent OF KED $t^{\text{OF}}$
for a replacement of $t^{\text{KS}}$ in a $t^{\text{KS}}$-MGGA xc-functional
with the hope of not changing much the results.
As mentioned above, a term $c\nabla^{2}\rho$
($c$ is a constant) in the KED [like in Eq.~(\ref{tGEA2L})] integrates to zero,
but would also not contribute to the kinetic potential
$\delta T_{\text{s}}/\delta\rho$ in a OF or embedding scheme since the contribution is
$\nabla^{2}\left(\partial\left(c\nabla^{2}\rho\right)/
\partial\left(\nabla^{2}\rho\right)\right)=\nabla^{2}c=0$.
However, MGGA xc-functionals depend on the KED in a more complicated way such that
$c\nabla^{2}\rho$ can not be discarded.

The $\nabla^{2}\rho$-dependent KED $t^{\text{OF}}$ that we will consider for a replacement of
$t^{\text{KS}}$ in xc-MGGAs are now listed (more detail can be found in the
respective references).
GEA2L\cite{KirzhnitsJETP57,BrackPLB76} as given by
Eq.~(\ref{tGEA2L}).
TW02L, which consists of the GGA TW02 proposed in
Ref.~\onlinecite{TranIJQC02} (a reparametrization of the
PBE GGA exchange\cite{PerdewPRL96} with $\kappa=0.8438$ and $\mu=0.2319$)
augmented with $\left(1/6\right)\nabla^{2}\rho$.
PC from Perdew and Constantin,\cite{PerdewPRB07} which was constructed
to recover the fourth-order GEA in the slowly varying
density limit and $t^{\text{W}}$ in the rapidly varying limit, as well as to
satisfy $t^{\text{W}}\leq t^{\text{PC}}$.
CR from Cancio and Redd\cite{CancioMP17} [Eqs.~(20) and (21) in
Ref.~\onlinecite{CancioMP17} with $\alpha=4$],
which was constructed in a rather similar way as PC.
GEAloc from Cancio and Redd\cite{CancioMP17}
[Eq.~(37) in Ref.~\onlinecite{CancioMP17}],
which has the same form as Eq.~(\ref{tGEA2L}) but with
different (optimized) parameters in front of $t^{\text{W}}$ and $\nabla^{2}\rho$.
PCopt and CRopt from Mejia-Rodriguez and Trickey\cite{MejiaRodriguezPRA17}
that are reoptimized versions of PC and CR, respectively.
Many other expressions for $t^{\text{OF}}$ could also be considered,
e.g., any of the integrand (augmented by $c\nabla^{2}\rho$)
of the numerous proposed $T_{s}^{\text{GGA}}$ or those proposed recently in
Refs.~\onlinecite{LindmaaPRB14,AstakhovIJQC16}.
Nevertheless, our selection of seven different OF KED should be good enough
to give us a general idea of the change in the performance of a xc-MGGA
when it is deorbitalized.

It is important to mention that for all considered OF KED, we chose to enforce
the lower bound $t^{\text{W}}\leq t$.\cite{HoffmannOstenhofPRA77,KurthIJQC99}
Thus, it is in fact
\begin{equation}
t^{\text{OF}'}(\mathbf{r})=
\max\left(t^{\text{OF}}(\mathbf{r}),t^{\text{W}}(\mathbf{r})\right)
\label{tOFmax}
\end{equation}
that replaces $t^{\text{KS}}$ in the MGGA xc-functionals,
which is also a way to locally reduce the error in $t^{\text{OF}}$.
Note that depending on the MGGA xc-functional,
Eq.~(\ref{tOFmax}) may be anyway necessary to apply if
negative values of $t^{\text{OF}}-t^{\text{W}}$ or $t^{\text{OF}}$ lead to
problems like, for instance, under a square root.

We also mention that the generalization of the OF KED formulas
for spin-polarized systems is trivially given by\cite{OliverPRA79}
$t[\rho_{\uparrow},\rho_{\downarrow}]=
t_{\uparrow}[\rho_{\uparrow}]+t_{\downarrow}[\rho_{\downarrow}]$, where
$t_{\sigma}[\rho_{\sigma}]=\left(1/2\right)t[2\rho_{\sigma}]$
with $t[2\rho_{\sigma}]$ being the non-spin-polarized formula in which
$\rho$ is replaced by $2\rho_{\sigma}$.

\subsection{\label{xcfunctionals}MGGA exchange-correlation functionals}

The MGGA xc-energy functionals that we will consider to test the accuracy
of OF KED are MVS\cite{SunPNAS15} and SCAN,\cite{SunPRL15} that
were used by Mejia-Rodriguez and
Trickey\cite{MejiaRodriguezPRA17} for their molecular tests,
as well as TM that was proposed by Tao and Mo.\cite{TaoPRL16}
The recent SCAN and TM functionals have been shown to be accurate
for many types of systems and properties
(see, e.g., Refs.~\onlinecite{TranJCP16,PengPRX16,MoJCP16,MoPRB17,HinumaPRB17}).
Additionally, the modified Becke-Johnson MGGA
potential\cite{TranPRL09} (mBJLDA, combined with LDA correlation\cite{PerdewPRB92a})
will also be used to test the accuracy of OF KED by considering the band gap.
The mBJLDA potential, which is based on the BJ potential,\cite{BeckeJCP06,TranJPCM07}
was shown to be much more reliable than the standard
LDA and GGA methods for band gap calculations and to lead to values
that are in very good agreement with experiment in most cases.
\cite{TranPRL09,SinghPRB10,JiangJCP13,TranJPCA17,NakanoJAP18,TranPRM18}

With an energy functional (MVS, SCAN, or TM), the closeness between
OF KEDs and the exact KS KED is quantified by considering properties that
depend on the total energy (lattice constant, bulk modulus, and cohesive
energy). With the mBJLDA potential, properties like band structure or
electron density are more interesting to look at.

\subsection{\label{detail}Computational details}

The calculations were done with WIEN2k,\cite{WIEN2k} which is an all-electron
code based on the linearized augmented plane-wave method.\cite{AndersenPRB75,Singh}
Very good parameters were chosen such that the results are well converged.
As in our previous work,\cite{TranJCP16} the lattice constant, bulk modulus, and
cohesive energy obtained with MGGAs were calculated using the GGA PBE\cite{PerdewPRL96}
orbitals and density since
in WIEN2k there is no implementation of the (non-multiplicative) potential
corresponding to a MGGA energy functional.
As discussed in Ref.~\onlinecite{TranJCP16}, the effect
of self-consistency on the results should be very small for strongly bound
(i.e., covalent, ionic, metallic) solids. However, self-consistency is expected
to affect more the results for weakly bound van der Waals solids.
Therefore, this is only via
the energy functional that the replacement $t^{\text{KS}}\rightarrow t^{\text{OF}}$
will produce changes in the lattice constant, bulk modulus, and cohesive energy.
The calculations of the band gap with the multiplicative mBJLDA potential
were done self-consistently.

\section{\label{results}Results}

\subsection{\label{lattice}Lattice constant, bulk modulus, and binding energy}

\begin{table*}[t]
\caption{\label{table_strong}The ME, MAE, MRE, and MARE of the parent
$t^{\text{KS}}$-MGGA functionals (MVS, SCAN, and TM) with respect to
experiment\cite{SchimkaJCP11,LejaeghereCRSSMS14} on the testing set of 44
strongly bound solids for the lattice constant $a_{0}$, bulk modulus $B_{0}$, and
cohesive energy $E_{\text{coh}}$.
The values for the $t^{\text{OF}}$-MGGA functionals are also with respect to
experiment, but with the value of the parent functional subtracted,
e.g., ME($t^{\text{OF}}$-MGGA)$-$ME($t^{\text{KS}}$-MGGA).
The units of the ME and MAE are \AA, GPa, and
eV/atom for $a_{0}$, $B_{0}$, and $E_{\text{coh}}$, respectively, and \% for the
MRE and MARE. The large differences with respect to the parent
$t^{\text{KS}}$-MGGA are underlined.
All results were obtained non-self-consistently using PBE orbitals/density.}
\begin{ruledtabular}
\begin{tabular}{lcccccccccccc}
\multicolumn{1}{l}{} &
\multicolumn{4}{c}{$a_ {0}$} &
\multicolumn{4}{c}{$B_ {0}$} &
\multicolumn{4}{c}{$E_{\text{coh}}$} \\
\cline{2-5}\cline{6-9}\cline{10-13}
\multicolumn{1}{l}{Functional} &
\multicolumn{1}{c}{ME} &
\multicolumn{1}{c}{MAE} &
\multicolumn{1}{c}{MRE} &
\multicolumn{1}{c}{MARE} &
\multicolumn{1}{c}{ME} &
\multicolumn{1}{c}{MAE} &
\multicolumn{1}{c}{MRE} &
\multicolumn{1}{c}{MARE} &
\multicolumn{1}{c}{ME} &
\multicolumn{1}{c}{MAE} &
\multicolumn{1}{c}{MRE} &
\multicolumn{1}{c}{MARE} \\
\hline
MVS          & -0.008             & 0.043             & -0.3             & 0.9             & 12.2              & 13.3             & 8.2               & 12.7             & 0.21              & 0.37              & 5.8               & 9.3              \\
\hline
MVS(GEA2L)   & \underline{-0.016} & -0.007            & \underline{-0.3} & -0.1            & -4.0              & -3.4             & -1.1              & -3.3             & -0.03             & -0.13             & -1.2              & -3.0             \\
MVS(TW02L)   & -0.007             & -0.009            & -0.1             & -0.2            & -4.7              & -3.6             & -2.5              & -4.0             & -0.13             & -0.13             & -3.9              & -2.6             \\
MVS(PC)      & -0.014             & -0.008            & -0.2             & -0.2            & -4.6              & -3.2             & -1.5              & -3.4             & -0.08             & -0.13             & -2.3              & -3.0             \\
MVS(CR)      & \underline{-0.016} & -0.007            & \underline{-0.3} & -0.1            & -3.9              & -3.4             & -1.1              & -3.3             & -0.02             & -0.12             & -0.8              & -2.9             \\
MVS(GEAloc)  & 0.006              & -0.007            & 0.2              & -0.1            & \underline{-9.3}  & \underline{-5.9} & -4.6              & \underline{-5.2} & \underline{-0.29} & -0.15             & \underline{-6.9}  & -3.4             \\
MVS(PCopt)   & -0.011             & 0.001             & -0.2             & 0.0             & \underline{-8.4}  & -3.8             & -3.0              & -3.2             & \underline{-0.25} & -0.08             & \underline{-5.3}  & -2.6             \\
MVS(CRopt)   & \underline{0.045}  & 0.007             & \underline{1.0}  & 0.1             & \underline{-17.1} & -3.2             & \underline{-11.8} & -3.7             & \underline{-0.59} & 0.07              & \underline{-14.1} & 1.4              \\
\hline
SCAN         & 0.018              & 0.030             & 0.3              & 0.6             & 3.5               & 7.4              & -0.5              & 6.5              & -0.02             & 0.19              & -0.7              & 4.9              \\
\hline
SCAN(GEA2L)  & -0.012             & -0.002            & -0.2             & 0.0             & -4.5              & 2.4              & -0.7              & 1.3              & 0.05              & -0.01             & 1.0               & -0.3             \\
SCAN(TW02L)  & -0.007             & -0.001            & -0.1             & 0.0             & \underline{-5.2}  & 2.5              & -1.6              & 1.5              & -0.00             & 0.00              & -0.5              & 0.1              \\
SCAN(PC)     & -0.010             & -0.001            & -0.2             & 0.0             & \underline{-5.0}  & 2.7              & -1.0              & 1.4              & 0.02              & 0.00              & 0.3               & 0.0              \\
SCAN(CR)     & -0.012             & -0.003            & -0.2             & 0.0             & -4.5              & 2.3              & -0.7              & 1.3              & 0.06              & -0.01             & 1.1               & -0.3             \\
SCAN(GEAloc) & \underline{0.016}  & 0.010             & \underline{0.4}  & 0.2             & \underline{-10.4} & 3.4              & -3.8              & 2.4              & \underline{-0.20} & 0.06              & -4.5              & 1.3              \\
SCAN(PCopt)  & -0.004             & -0.002            & 0.0              & 0.0             & \underline{-6.4}  & 0.3              & -1.8              & 0.2              & -0.07             & -0.02             & -1.6              & -0.1             \\
SCAN(CRopt)  & \underline{0.034}  & \underline{0.023} & \underline{0.8}  & \underline{0.5} & \underline{-11.7} & 3.8              & \underline{-6.2}  & 3.4              & \underline{-0.28} & 0.12              & \underline{-6.6}  & 3.0              \\
\hline
TM           & -0.006             & 0.023             & -0.2             & 0.5             & 2.4               & 6.6              & 2.1               & 6.2              & 0.24              & 0.27              & 6.4               & 7.0              \\
\hline
TM(GEA2L)    & -0.005             & 0.002             & -0.1             & 0.0             & -0.9              & 0.9              & -0.5              & 0.4              & -0.01             & 0.01              & -0.3              & 0.2              \\
TM(TW02L)    & -0.003             & 0.001             & -0.1             & 0.0             & -0.9              & 0.9              & -0.8              & 0.3              & -0.02             & 0.01              & -0.7              & 0.0              \\
TM(PC)       & -0.006             & 0.003             & -0.1             & 0.1             & -0.7              & 1.0              & -0.1              & 0.6              & -0.02             & 0.02              & -0.5              & 0.4              \\
TM(CR)       & -0.005             & 0.002             & -0.1             & 0.0             & -0.8              & 0.9              & -0.5              & 0.4              & -0.00             & 0.01              & -0.1              & 0.2              \\
TM(GEAloc)   & -0.010             & 0.003             & -0.2             & 0.1             & -0.4              & 1.6              & 0.9               & 1.2              & -0.01             & 0.03              & 0.2               & 1.1              \\
TM(PCopt)    & 0.004              & 0.004             & 0.1              & 0.1             & -2.9              & 1.7              & -1.5              & 0.9              & -0.09             & -0.03             & -2.0              & -0.6             \\
TM(CRopt)    & 0.009              & 0.004             & 0.2              & 0.1             & -3.6              & 1.0              & -2.4              & 0.5              & \underline{-0.13} & -0.05             & -2.8              & -1.0             \\
\end{tabular}
\end{ruledtabular}
\end{table*}

\begin{figure}
\includegraphics[scale=0.4]{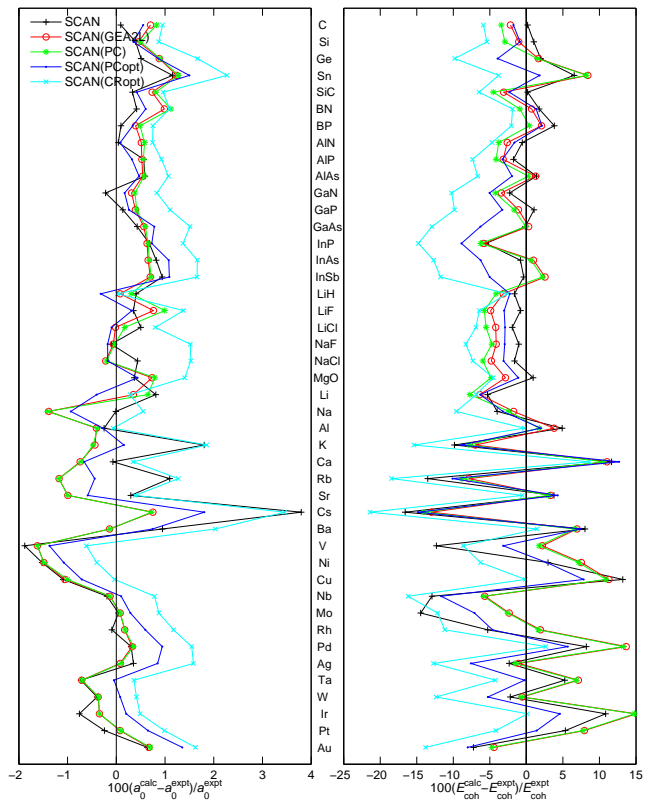}
\caption{\label{fig_solids_1}Relative error (in \%) with respect to
experiment\cite{SchimkaJCP11,LejaeghereCRSSMS14} in the calculated lattice
constant (left panel) and cohesive energy (right panel) for the 44 strongly
bound solids.}
\end{figure}

\begin{table}
\caption{\label{results_RG}Equilibrium lattice constant $a_{0}$ (in \AA) and
cohesive energy $E_{\text{coh}}$ (in meV/atom) of rare-gas solids. 
The values for the $t^{\text{OF}}$-MGGA functionals
are the difference from those obtained with the parent $t^{\text{KS}}$-MGGA,
e.g., $a_{0}(t^{\text{OF}}\text{-MGGA})-a_{0}(t^{\text{KS}}\text{-MGGA})$.
The reference CCSD(T) results, which agree closely with experiment,\cite{RosciszewskiPRB00}
are also shown. The large differences with respect to the parent
$t^{\text{KS}}$-MGGA are underlined.
All results were obtained non-self-consistently using the PBE orbitals/density.}
\begin{ruledtabular}
\begin{tabular}{lcccccc}
\multicolumn{1}{l}{} &
\multicolumn{2}{c}{Ne} &
\multicolumn{2}{c}{Ar} &
\multicolumn{2}{c}{Kr} \\
\cline{2-3}\cline{4-5}\cline{6-7}
\multicolumn{1}{l}{Method} &
\multicolumn{1}{c}{$a_{0}$} &
\multicolumn{1}{c}{$E_{\text{coh}}$} &
\multicolumn{1}{c}{$a_{0}$} &
\multicolumn{1}{c}{$E_{\text{coh}}$} &
\multicolumn{1}{c}{$a_{0}$} &
\multicolumn{1}{c}{$E_{\text{coh}}$} \\
\hline
MVS          &  4.02             &  59             &  5.41 &  56 &  5.79 &     69  \\
\hline
MVS(GEA2L)   & \underline{-0.14} &  \underline{41} & \underline{-0.34} &  \underline{70} & \underline{-0.30} & \underline{80}  \\
MVS(TW02L)   & -0.03             &   0             & \underline{-0.21} &  \underline{29} & \underline{-0.17} & \underline{33}  \\
MVS(PC)      & \underline{-0.15} &  \underline{47} & \underline{-0.34} &  \underline{75} & \underline{-0.31} & \underline{85}  \\
MVS(CR)      & \underline{-0.14} &  \underline{41} & \underline{-0.34} &  \underline{70} & \underline{-0.30} & \underline{80}  \\
MVS(GEAloc)  & \underline{-0.11} &  \underline{31} & \underline{-0.26} &  \underline{54} & \underline{-0.19} & \underline{55}  \\
MVS(PCopt)   & \underline{-0.13} &  \underline{38} & \underline{-0.31} &  \underline{66} & \underline{-0.23} & \underline{63}  \\
MVS(CRopt)   & \underline{0.85}  & \underline{-53} & \underline{1.03}  & \underline{-48} &  \underline{0.75} & \underline{-53} \\
\hline
SCAN         &  4.03             &  54             &  5.31             &  61             &  5.74             & 72              \\
\hline
SCAN(GEA2L)  & -0.02             &  \underline{11} & -0.15             &  \underline{32} & \underline{-0.20} & \underline{50}  \\
SCAN(TW02L)  &  0.03             &  \underline{-6} & -0.08             &   9             & -0.15             & 23              \\
SCAN(PC)     & -0.03             &  \underline{15} & -0.15             &  \underline{36} & \underline{-0.20} & \underline{54}  \\
SCAN(CR)     & -0.02             &  \underline{12} & -0.15             &  \underline{32} & \underline{-0.20} & \underline{50}  \\
SCAN(GEAloc) &  0.02             &   5             & -0.06             &  19             & -0.11             & \underline{33}  \\
SCAN(PCopt)  & -0.03             &  \underline{12} & -0.11             &  \underline{28} & -0.14             & \underline{40}  \\
SCAN(CRopt)  & \underline{0.63}  & \underline{-48} &  \underline{0.25} & \underline{-37} &  \underline{0.26} & \underline{-37} \\
\hline
TM           &  4.05             &  47             &  5.23             &  62             &  5.60             & 82              \\
\hline
TM(GEA2L)    & -0.00             &   \underline{7} & -0.08             &  \underline{22} & -0.08             & \underline{27}  \\
TM(TW02L)    &  0.03             &  -5             & -0.05             &   9             & -0.05             & 13              \\
TM(PC)       & -0.03             &  \underline{-8} & -0.14             &  12             & -0.14             & 22              \\
TM(CR)       & -0.00             &   \underline{7} & -0.08             &  \underline{22} & -0.08             & \underline{27}  \\
TM(GEAloc)   & -0.10             &  \underline{32} & \underline{-0.17} &  \underline{56} & \underline{-0.16} & \underline{67}  \\
TM(PCopt)    & -0.01             & \underline{-10} & -0.11             &   7             & -0.11             & 15              \\
TM(CRopt)    &  0.05             &  -4             & -0.01             &   7             & -0.01             & 9               \\
\hline
Reference     &  4.30             &  26             &  5.25 &  88 &  5.60 &    122 \\
\end{tabular}
\end{ruledtabular}
\end{table}

\begin{table*}
\caption{\label{results_LC}Equilibrium lattice constant $c_{0}$ (in \AA) and
interlayer binding energy $E_{\text{b}}$ (in meV/atom) of layered solids.
The values for the $t^{\text{OF}}$-MGGA functionals are the difference
from those obtained with the parent $t^{\text{KS}}$-MGGA, e.g.,
$c_{0}(t^{\text{OF}}\text{-MGGA})-c_{0}(t^{\text{KS}}\text{-MGGA})$.
The intralayer constant $a$
was not optimized, but kept fixed at the experimental value.\cite{BjorkmanPRB12}
Reference results\cite{BjorkmanPRB12} from experiment for $c_{0}$ and from RPA
for $E_{\text{b}}$ are also shown. The large differences with respect to the parent
$t^{\text{KS}}$-MGGA are underlined.
All results were obtained non-self-consistently using PBE
orbitals/density.}
\begin{ruledtabular}
\begin{tabular}{lcccccccccc}
\multicolumn{1}{l}{} &
\multicolumn{2}{c}{Graphite} &
\multicolumn{2}{c}{h-BN} &
\multicolumn{2}{c}{TiS$_{2}$} &
\multicolumn{2}{c}{MoTe$_{2}$} &
\multicolumn{2}{c}{WSe$_{2}$} \\
\cline{2-3}\cline{4-5}\cline{6-7}\cline{8-9}\cline{10-11}
\multicolumn{1}{l}{Method} &
\multicolumn{1}{c}{$c_{0}$} &
\multicolumn{1}{c}{$E_{\text{b}}$} &
\multicolumn{1}{c}{$c_{0}$} &
\multicolumn{1}{c}{$E_{\text{b}}$} &
\multicolumn{1}{c}{$c_{0}$} &
\multicolumn{1}{c}{$E_{\text{b}}$} &
\multicolumn{1}{c}{$c_{0}$} &
\multicolumn{1}{c}{$E_{\text{b}}$} &
\multicolumn{1}{c}{$c_{0}$} &
\multicolumn{1}{c}{$E_{\text{b}}$} \\
\hline
MVS          &  6.60 & 32 &  6.43 & 38 &  5.79 &  30 & 14.66 &  34 & 13.48 & 19 \\
\hline
MVS(GEA2L)   & \underline{-0.24} & \underline{13} & \underline{-0.21} & 10             & \underline{-0.19}&  \underline{18} & \underline{-0.25} &   6             & \underline{-0.22} & \underline{12} \\
MVS(TW02L)   & \underline{-0.22} & \underline{11} & \underline{-0.19} &  8             & \underline{-0.12}&   9             & -0.13             &   0             & -0.09             &  5 \\
MVS(PC)      & \underline{-0.24} & \underline{13} & \underline{-0.20} & 10             & \underline{-0.14}&  \underline{17} & \underline{-0.25} &   7             & \underline{-0.22} & \underline{12} \\
MVS(CR)      & \underline{-0.24} & \underline{13} & \underline{-0.21} & 10             & \underline{-0.19}&  \underline{18} & \underline{-0.25} &   6             & \underline{-0.22} & \underline{12} \\
MVS(GEAloc)  & \underline{-0.14} & 10             & \underline{-0.13} &  7             &  0.02            &   7             &  0.12             &  -2             &  0.07             &  4 \\
MVS(PCopt)   & \underline{-0.13} & 10             & \underline{-0.12} &  7             & -0.07            &  \underline{11} &  0.01             &   2             &  0.02             &  7 \\
MVS(CRopt)   &  0.02             & -1             &  \underline{0.14} & -7             & \underline{0.28} & \underline{-11} &  \underline{0.37} & \underline{-13} &  \underline{0.59} & -8 \\
\hline
SCAN         &  6.94             & 20             &  6.79             & 21             &  5.93            &  21             & 14.75             &  30             & 13.68             & 17 \\
\hline
SCAN(GEA2L)  & \underline{-0.13} &  4             & -0.10             &  5             & \underline{-0.12}&  \underline{12} & \underline{-0.33} &   8             &\underline{-0.26}  & 10 \\
SCAN(TW02L)  & -0.10             &  2             & -0.08             &  3             & -0.09            &   8             & \underline{-0.31} &   5             & \underline{-0.23} &  7 \\
SCAN(PC)     & \underline{-0.13} &  4             & -0.10             &  5             & -0.09            &  \underline{11} & \underline{-0.33} &   8             & \underline{-0.26} & 10 \\
SCAN(CR)     & \underline{-0.13} &  4             & -0.10             &  5             & \underline{-0.12}&  \underline{12} & \underline{-0.33} &   8             & \underline{-0.26} & 10 \\
SCAN(GEAloc) & -0.09             &  3             & -0.06             &  4             &  0.03            &   5             &  0.13             &  -1             &  0.05             &  3 \\
SCAN(PCopt)  & \underline{-0.12} &  3             & -0.08             &  4             & -0.04            &   8             &  0.07             &   0             &  0.01             &  4 \\
SCAN(CRopt)  &  0.03             & -1             &  0.05             & -2             &  \underline{0.16}&  -5             &  \underline{0.41} &  -9             & \underline{0.36}  & -5 \\
\hline
TM           &  6.63             & 29             &  6.49             & 32             &  5.73            &  44             & 14.17             &  50             & 13.21             & 35 \\
\hline
TM(GEA2L)    & -0.08             &  4             & -0.06             &  3             & -0.08            &   7             & -0.16             &   7             & -0.11             &  6 \\
TM(TW02L)    & -0.09             &  4             & -0.07             &  3             & -0.08            &   6             & -0.16             &   7             & -0.11             &  5 \\
TM(PC)       & \underline{-0.15} &  4             & \underline{-0.11} &  4             & -0.07            &   6             & -0.17             &   8             & -0.14             &  6 \\
TM(CR)       & -0.08             &  4             & -0.06             &  3             & -0.08            &   7             & -0.16             &   7             & -0.11             &  6 \\
TM(GEAloc)   & \underline{-0.23} & \underline{17} & \underline{-0.21} & \underline{15} & -0.07            &  \underline{15} & -0.14             &  \underline{14} & -0.15             & \underline{13} \\
TM(PCopt)    & \underline{-0.12} &  3             & -0.08             &  2             &  0.02            &   3             &  0.02             &   2             &  0.02             &  2 \\
TM(CRopt)    & -0.10             &  7             & -0.07             &  5             &  0.02            &   3             &  0.05             &   2             &  0.03             &  2 \\
\hline
Reference     &  6.70             & 48             &  6.69             & 40             &  5.71            &  95             & 13.97             & 111             & 12.96             & 93 \\
\end{tabular}
\end{ruledtabular}
\end{table*}

We start with the results for the equilibrium lattice constant $a_{0}$,
bulk modulus $B_{0}$, and cohesive energy $E_{\text{coh}}$ of 44 strongly
bound solids (listed in Table~S1 of the supplementary material\cite{SM_tau}).
Table~\ref{table_strong} shows the mean error (ME),
mean absolute error (MAE), mean relative error (MRE), and mean absolute
relative error (MARE) with respect to experiment. The values of $a_{0}$,
$B_{0}$, and $E_{\text{coh}}$ can be found in Tables~S1-S9
and Figs.~S1-S24 of the supplementary material.\cite{SM_tau}
The errors obtained with the parent $t^{\text{KS}}$-MGGA, namely, MVS, SCAN, or TM,
are considered as the reference that should be reproduced at best by an OF
$t^{\text{OF}}$-MGGA [denoted MGGA(X), where X is one of the OF
approximations $t^{\text{OF}}$ mentioned in Sec.~\ref{kinetic}].
Since the amount of results shown in Table~\ref{table_strong} is rather
substantial and would make a detailed discussion rather lengthy and tedious,
a concise discussion, only in terms of MAE and ME, of the most
interesting observations is provided.

In the case of the SCAN and TM xc-functionals, the deorbitalization
procedure leads to changes in the MAE and ME that are the smallest if
$t^{\text{KS}}$ is replaced by $t^{\text{GEA2L}}$, $t^{\text{TW02L}}$,
$t^{\text{PC}}$, or $t^{\text{CR}}$. The change in the MAE is in most cases
below $0.003$~\AA~for $a_{0}$, 2.5~GPa for $B_{0}$, and 0.03~eV/atom for
$E_{\text{coh}}$, such that it is reasonable to consider the overall
(in)accuracy of the xc-functional as unaffected by its deorbitalization.
$t^{\text{PCopt}}$ also belongs to the group of the accurate OF KED in the case of SCAN,
but not TM especially for the bulk modulus and cohesive energy.
If the deorbitalization of SCAN or TM is done with
$t^{\text{GEAloc}}$ or $t^{\text{CRopt}}$,
then larger changes in the MAE and ME can sometimes,
but not systematically, be observed. This seems to be particularly
the case with $t^{\text{CRopt}}$, which, for instance, leads for SCAN
to changes of 0.023~\AA~and 3.8~GPa in the MAE of $a_{0}$ and $B_{0}$
respectively. $t^{\text{CRopt}}$ also leads to the largest change in the
MAE of $a_{0}$ and $E_{\text{coh}}$ for TM.
Thus, replacing $t^{\text{KS}}$
by $t^{\text{GEAloc}}$ or $t^{\text{CRopt}}$, in particular,
affects more the accuracy of a functional and would probably change the
position of the xc-functional in some performance ranking
(see Ref.~\onlinecite{TranJCP16}).

Compared to SCAN and TM, the deorbitalization procedure of MVS leads
to changes in the MAE that are in general clearly larger.
This is due to the analytical form of MVS
that depends more strongly on the KED.
For instance, for $B_{0}$ there is a decrease in the MAE that
is in the range 3.2-5.9~GPa, while for $E_{\text{coh}}$ the MAE of the
$t^{\text{OF}}$-MVS can be decreased by 0.15~eV/atom [with MVS(GEAloc)]
or increased by 0.07~eV/atom [with MVS(CRopt)].
Concerning the ME of MVS, $t^{\text{GEA2L}}$, $t^{\text{TW02L}}$,
$t^{\text{PC}}$, and $t^{\text{CR}}$ are more
efficient than $t^{\text{GEAloc}}$, $t^{\text{PCopt}}$, and $t^{\text{CRopt}}$
for reproducing the values of MVS. Note that in terms of MAE, MVS(CRopt)
seems to be the closest to MVS, but this is fortuitous since
the ME are completely different and of opposite sign.

Figure~\ref{fig_solids_1} shows for each solid the relative error in the
lattice constant and cohesive energy obtained with the parent SCAN and four
of its deorbitalized versions. We can see that the results with
SCAN(GEA2L) and SCAN(PC), which are basically the same,
are very or fairly close to SCAN results in most cases.
The most visible exceptions are the alkali and alkaline earth metals for
which the SCAN(CRopt) values follow very closely those obtained with SCAN
in particular for $a_{0}$. We also note some large differences
in $E_{\text{coh}}$ between SCAN(GEA2L/PC) and SCAN for
some of the $3d$ and $4d$ transition metals and the ionic compounds.
Except for the aforementioned alkali and alkaline earth metals,
the lattice constants and cohesive energies obtained with SCAN(CRopt)
differ noticeably from SCAN. SCAN(PCopt) leads to results that
are intermediate between SCAN(GEA2L/PC) and SCAN(CRopt).

Thus, in summary the performance of a xc-MGGA functional for strongly bound
solids is modified the least when $t^{\text{KS}}$ is replaced by
$t^{\text{GEA2L}}$, $t^{\text{TW02L}}$, $t^{\text{PC}}$, $t^{\text{CR}}$, or
$t^{\text{PCopt}}$.
For SCAN and TM the performance is overall little affected by the
deorbitalization using one these OF KED, but more for MVS.

Although the goal of replacing $t^{\text{KS}}$ by $t^{\text{OF}}$ in a
xc-MGGA was not to improve the agreement with experiment, we mention that 
it is sometimes the case.
By looking at the MA(R)E in Table~\ref{table_strong},
we can see that, for instance, the deorbitalizion of MVS reduces the values
for $a_{0}$, $B_{0}$, and $E_{\text{coh}}$.

In their work, Mejia-Rodriguez and Trickey\cite{MejiaRodriguezPRA17}
reported changes (due to the deorbitalization) in the MAE for bond lengths
of molecules that are below 0.002~\AA~with MVS, which is small.
The change in the ME can be larger in some cases, since while the ME is
$-0.0016$~\AA~with MVS, it increases to 0.0069~\AA~with MVS(PC), but
is rather similar, $-0.0025$~\AA, with MVS(PCopt).
The deorbitalisation of SCAN leads to larger changes in the MAE
of bond lengths (up to $\sim0.01$~\AA), but not for the ME since the largest
change is $\sim0.016$~\AA, which is barely larger than for MVS.
From these results on molecular bond lengths, $t^{\text{PCopt}}$ seems to be
a more accurate OF KED than the others. This is in line with our observation
that $t^{\text{PCopt}}$ is among the most accurate OF KED for
the lattice constants of solids.
Concerning the heat of formation,\cite{MejiaRodriguezPRA17}
the changes in the MAE and ME seem to be in many cases the smallest
with $t^{\text{PCopt}}$, as well. For instance, the deorbitalization of SCAN leads
to a change in the MAE of $+15$ and $+0.5$~kcal/mol with
$t^{\text{PC}}$ and $t^{\text{PCopt}}$, respectively,
and $+21$ and $+6$~kcal/mol for the ME.
We also mention that from the results of Mejia-Rodriguez and Trickey, we can not
observe a change in the results due to the deorbitalization that is larger
in the case of MVS as we did.

Turning now to weakly bound van der Waals systems, Tables~\ref{results_RG} and
\ref{results_LC} show the results for rare-gas (Ne, Ar, and Kr)
and layered hexagonal solids (graphite, h-BN, TiS$_{2}$, MoTe$_{2}$, and WSe$_{2}$),
respectively.

The range of errors in the lattice constant obtained in typical performance
tests of DFT functionals on van der Waals systems
(see, e.g., Refs.~\onlinecite{JohnsonCPL04,ZhaoJPCA06,TranJCP13,RegoJPCM15,TranJCP16})
is by far much larger than for covalent or ionic solids.
Hence, for our systems it should be fair to consider that the performance of
a $t^{\text{KS}}$-MGGA (with respect to other functionals) is not really
modified by its deorbitalization if the change in the lattice constant is,
let us say, below something like (this may be a matter of personal taste)
$\sim0.1$-$0.15$~\AA~for the rare-gas ($a_{0}$) and layered solids ($c_{0}$).
With this criterion, the results show that the replacement
$t^{\text{KS}}\rightarrow t^{\text{OF}}$ in SCAN and TM leads to acceptable
changes in the lattice constant in most cases except maybe Kr.
With MVS, however, the changes are two or three times larger and
unacceptable since they may
affect the performance of MVS with respect to other functionals.

By choosing, again arbitrarily, $\sim20\%$ of the reference
coupled cluster [CCSD(T)] or random-phase approximation (RPA)
binding energy as the
largest change that can be accepted when a functional is deorbitalized,
then too large variations in $E_{\text{coh}}$ or $E_{\text{b}}$
are usually observed for MVS, especially for the rare gases.
The deorbitalization of SCAN or TM affects less the results, but
nevertheless the change for the rare gases is in most cases also
too large according to our criterion. Interestingly, note that the
deorbitalization of the SCAN and TM functionals leads in many cases
to a better agreement with CCSD(T) for the binding energy.

For the rare gases, the OF KED that leads overall to the smallest
perturbations for the deorbitalization of the xc-MGGAs seems to be $t^{\text{TW02L}}$.
Note that $t^{\text{CRopt}}$ shows rather strange results
since it is the worst when used in MVS and SCAN, while it is the best for TM. 
In the case of the layered solids, a good choice for $t^{\text{OF}}$
is $t^{\text{GEAloc}}$ for MVS and SCAN,
while with TM all $t^{\text{OF}}$ except $t^{\text{GEAloc}}$
are of similar accuracy.

\subsection{\label{bandgap}Band gaps}

\begin{table}
\caption{\label{table_bandgap}The ME, MAE, MRE, and MARE (with respect to
experiment
\cite{CrowleyJPCL16,LuceroJPCM12,BernstorffOC86,GillenJPCM13,SchimkaJCP11,KollerJPCM13,SkonePRB14,ShiPRB05,LeePRB16,GanoseJMCC16,GrohJPCS09})
on the testing set of 76 solids
(listed in Table~S10 of the supplementary material\cite{SM_tau})
for the fundamental band gap $E_{\text{g}}$
obtained with mBJLDA and its deorbitalized versions, as well as PBE and HSE06.
The units are eV for the ME and MAE and \% for the MRE and MARE.}
\begin{ruledtabular}
\begin{tabular}{lcccc} 
                & ME    & MAE  & MRE & MARE \\
\hline
mBJLDA         & -0.30 & 0.47 &  -5 & 15   \\
mBJLDA(GEA2L)  & -0.95 & 0.97 & -32 & 32   \\
mBJLDA(TW02L)  & -1.03 & 1.03 & -33 & 33   \\
mBJLDA(PC)     & -1.17 & 1.18 & -32 & 33   \\
mBJLDA(CR)     & -0.94 & 0.96 & -31 & 32   \\
mBJLDA(GEAloc) &  0.39 & 0.92 &   6 & 21   \\
mBJLDA(PCopt)  & -0.54 & 0.67 & -10 & 16   \\
mBJLDA(CRopt)  & -0.08 & 0.75 & -10 & 19   \\
PBE             & -1.99 & 1.99 & -53 & 53   \\
HSE06           & -0.68 & 0.82 &  -7 & 17   \\
\end{tabular} 
\end{ruledtabular}
\end{table} 

\begin{figure}
\includegraphics[width=\columnwidth]{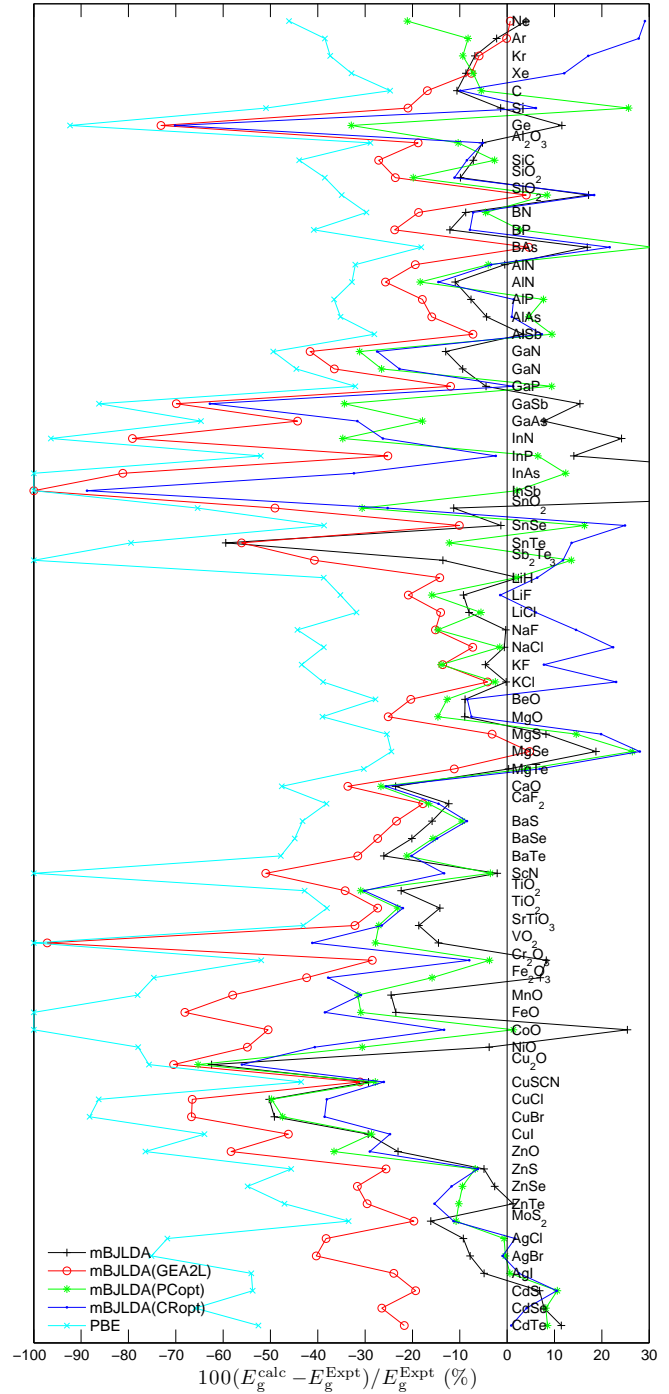}
\caption{\label{fig_band_gap_ref}Relative error (in \%) with respect to
experiment
\cite{CrowleyJPCL16,LuceroJPCM12,BernstorffOC86,GillenJPCM13,SchimkaJCP11,KollerJPCM13,SkonePRB14,ShiPRB05,LeePRB16,GanoseJMCC16,GrohJPCS09}
in the band gap $E_{\text{g}}$.}
\end{figure}

Turning to the electronic structure, Table~\ref{table_bandgap} and
Fig.~\ref{fig_band_gap_ref} (for selected methods) show the
results obtained for the fundamental band gap $E_{\text{g}}$ calculated with the mBJLDA
potential and its deorbitalized versions. The testing set, which was used in
our previous works,\cite{TranJPCA17,TranPRM18} consists of 76 solids
(listed in Table~S10 of the supplementary material\cite{SM_tau}) of
various types: ionic insulators, $sp$-semiconductors, rare gases, as well
as strongly correlated solids.
As shown in Refs.~\onlinecite{TranJPCA17,TranPRM18}, the mBJLDA potential is
on average more accurate for the band gap than all other semilocal potentials and
hybrid functionals that were considered for comparison
(the PBE\cite{PerdewPRL96} and HSE06\cite{HeydJCP03,KrukauJCP06} results
are also shown in Table~\ref{table_bandgap} and Fig.~\ref{fig_band_gap_ref}).

From the statistics shown in Table~\ref{table_bandgap}, the first observation
is that deorbitalizing the mBJLDA potential leads to an increase of
the MAE and MARE, no matter what OF KED is used. The deterioration
is the smallest when $t^{\text{KS}}$ is replaced by $t^{\text{PCopt}}$, and in this
case the MAE increases from 0.47 to 0.67~eV and the MARE from 15 to 16\%.
This increase in the MARE is clearly
negligible, but also quite acceptable for the MAE considering that most
other potentials lead to larger MAE for this test set.\cite{TranJPCA17,TranPRM18}
With mBJLDA(CRopt), a small increase of 4\% for the MARE is obtained, while the
MAE increases to 0.75~eV, which is now on the verge of being acceptable since
other potentials, e.g., AK13,\cite{ArmientoPRL13} B3PW91,\cite{BeckeJCP93b}
or HSE06\cite{HeydJCP03,KrukauJCP06}
lead to similar MAE.\cite{TranJPCA17,TranPRM18}
Substituting $t^{\text{KS}}$ by any of the other OF KED leads to a clearly
larger MAE (around 1~eV) and MARE (above 30\%, except with $t^{\text{GEAloc}}$).

Looking into more detail at the results (see Table~S11 and Figs.~S25-S32 of
the supplementary material\cite{SM_tau} and Fig.~\ref{fig_band_gap_ref}),
we can see that an inaccurate OF KED like $t^{\text{GEA2L}}$ leads
to band gaps which are in most cases about halfway between the mBJLDA
and PBE values, such that a rather clear underestimation is obtained
on average (see ME and MRE in Table~\ref{table_bandgap}).
The mBJLDA band gaps are in general reproduced more accurately by
mBJLDA(PCopt) and/or mBJLDA(CRopt) except for the rare gases for
which mBJLDA(GEA2L) is the closest to mBJLDA.

Finally, we note that a reoptimization of the parameters $\alpha$ and $\beta$
in a OF mBJLDA potential [see Ref.~\onlinecite{TranPRL09} for details] may
possibly lead to a (partial) recovery of the performance of the original mBJLDA
potential. However, we have not made any attempts since this is beyond the scope
of this work.

\section{\label{discussion}Further Discussion}

Thanks to their additional dependency on $t^{\text{KS}}$, $t^{\text{KS}}$-MGGAs
are more flexible than GGAs and, therefore, have the possibility to be
universally more accurate. As shown above,
a $t^{\text{KS}}$-MGGA can be replaced rather
efficiently (albeit not systematically) by a corresponding $\nabla^{2}\rho$-MGGA, and
in order to shed some light on the relation between
$t^{\text{KS}}$ and $\nabla^2\rho$,
a principal component analysis\cite{PearsonPM01,Jolliffe}
(PCA) of $t^{\text{TF}}$, $t^{\text{W}}$,
$\nabla^2\rho$, and $t^{\text{KS}}$ has been carried out.
From the PCA, an approximation for $t^{\text{KS}}$ that consists of
a linear combination of $t^{\text{TF}}$, $t^{\text{W}}$, and $\nabla^2\rho$
is obtained, and its accuracy reveals to which extent
$t^{\text{KS}}$ can be represented by $\rho$ and its first two derivatives.

The $4\times4$ covariance matrix was calculated using uniformly sampled data from one
representative of metallic (Cu), layered (graphite), and covalently bound (Si)
systems, and diagonalized in order to get the eigenvalues and
corresponding eigenvectors spanning the four-dimensional space of
$t^{\text{TF}}$, $t^{\text{W}}$, $\nabla^2\rho$, and
$t^{\text{KS}}$. In the next step, we neglect the eigenvector with
the smallest eigenvalue, thereby obtaining the three dimensional
representation which explains most of the variance in the data.
Now, assuming that all points are on this three dimensional hyperplane,
one can reconstruct an OF KED from the values of
$\rho$ (i.e., $t^{\text{TF}}$), $\nabla\rho$ (i.e., $t^{\text{W}}$),
and $\nabla^2\rho$, and the resulting linear combination is given by
\begin{equation}
t^{\text{PCA}}(\mathbf{r})=1.069t^{\text{TF}}(\mathbf{r}) -
0.244t^{\text{W}}(\mathbf{r})+0.438\nabla^2\rho(\mathbf{r}).
\end{equation}
The coefficient in front of $t^{\text{TF}}$ is close to 1 as it should be
in order to recover the homogeneous electron gas limit,
while those in front of $t^{\text{W}}$ and
$\nabla^2\rho$ show big differences from GEA2L [Eq.~(\ref{tGEA2L})].
However, it is worth
mentioning that a negative coefficient in front of $t^{\text{W}}$ is found
also in GEAloc\cite{CancioMP17} ($-0.165$) and in a KED expression derived
for the Airy gas\cite{LindmaaPRB14} ($-1/9\approx-0.111$).

\begin{figure}
\includegraphics[width=1.0\columnwidth]{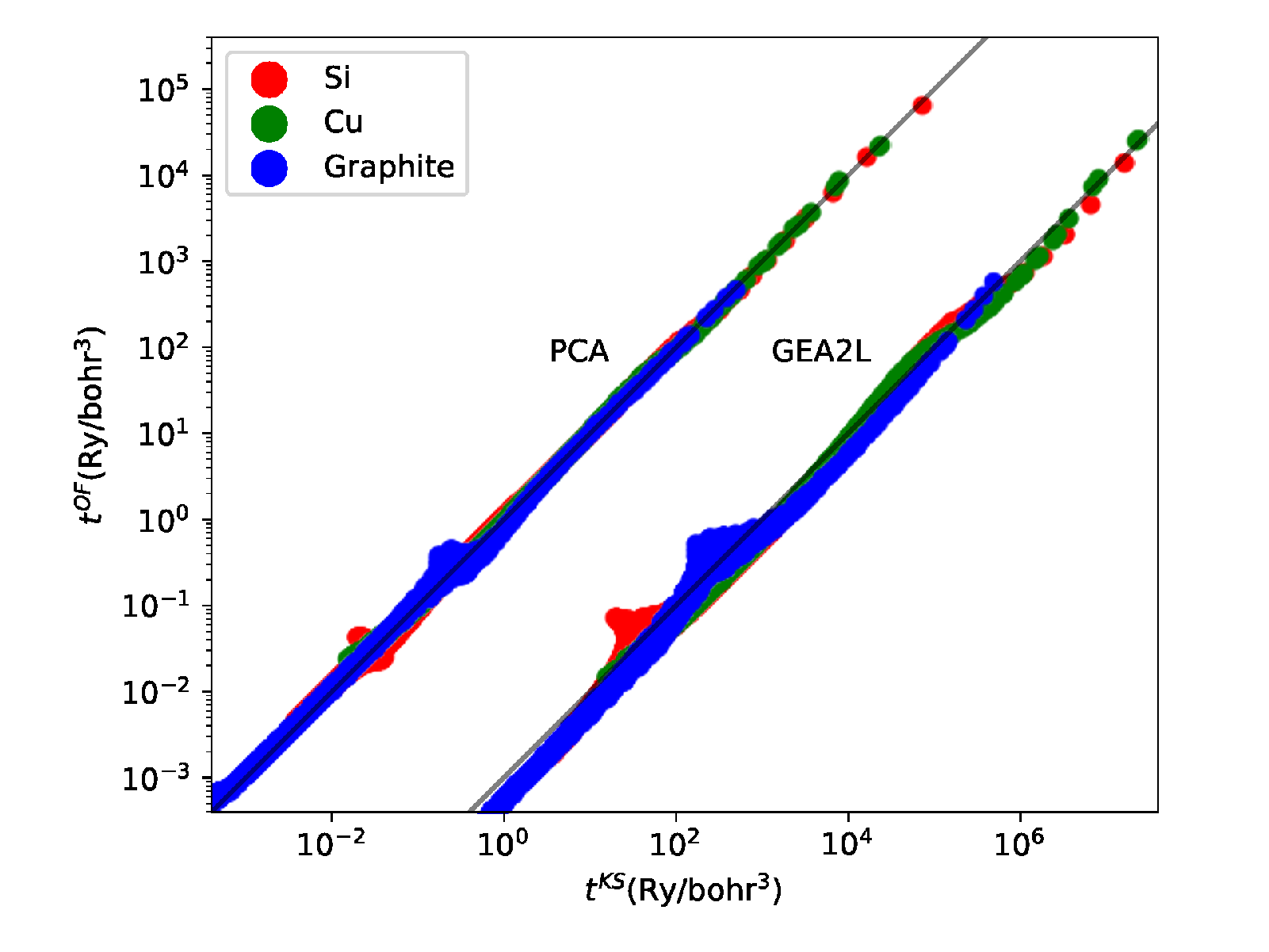}
\caption{Comparison between the KS KED and the GEA2L and PCA approximations
for different solids. For clarity (no overlap between the GEA2L and PCA data),
the $t^{\text{KS}}$ values for GEA2L are multiplied by 1000 (i.e., right shifted).
A perfect approximation should coincide with the diagonal solid black line.}
\label{fig:all}
\end{figure}
Figure~\ref{fig:all} shows for the three selected solids the accuracy of the
GEA2L and our PCA approximation with the Weizs\"{a}cker lower bound
enforced [Eq.~(\ref{tOFmax})].
We can see that the PCA approximation shows better agreement with
the KS KED than GEA2L, similarly as obtained by Seino \textit{et al}.
\cite{SeinoJCP18} for atoms and small organic molecules using a
machine learning algorithm. It is also
important to note that for both approximations there are two regions where one
can find larger errors. These two lumps are from Si and graphite,
where GEA2L systematically overestimates the KED, while in the PCA approximation
these errors are still there but largely reduced. Actually, the errors for graphite can be
found in the same KED region as the errors for organic molecules.\cite{SeinoJCP18}

\begin{figure}
\begin{picture}(7,14)(0,0)
\put(0,8){\epsfxsize=7cm \epsfbox{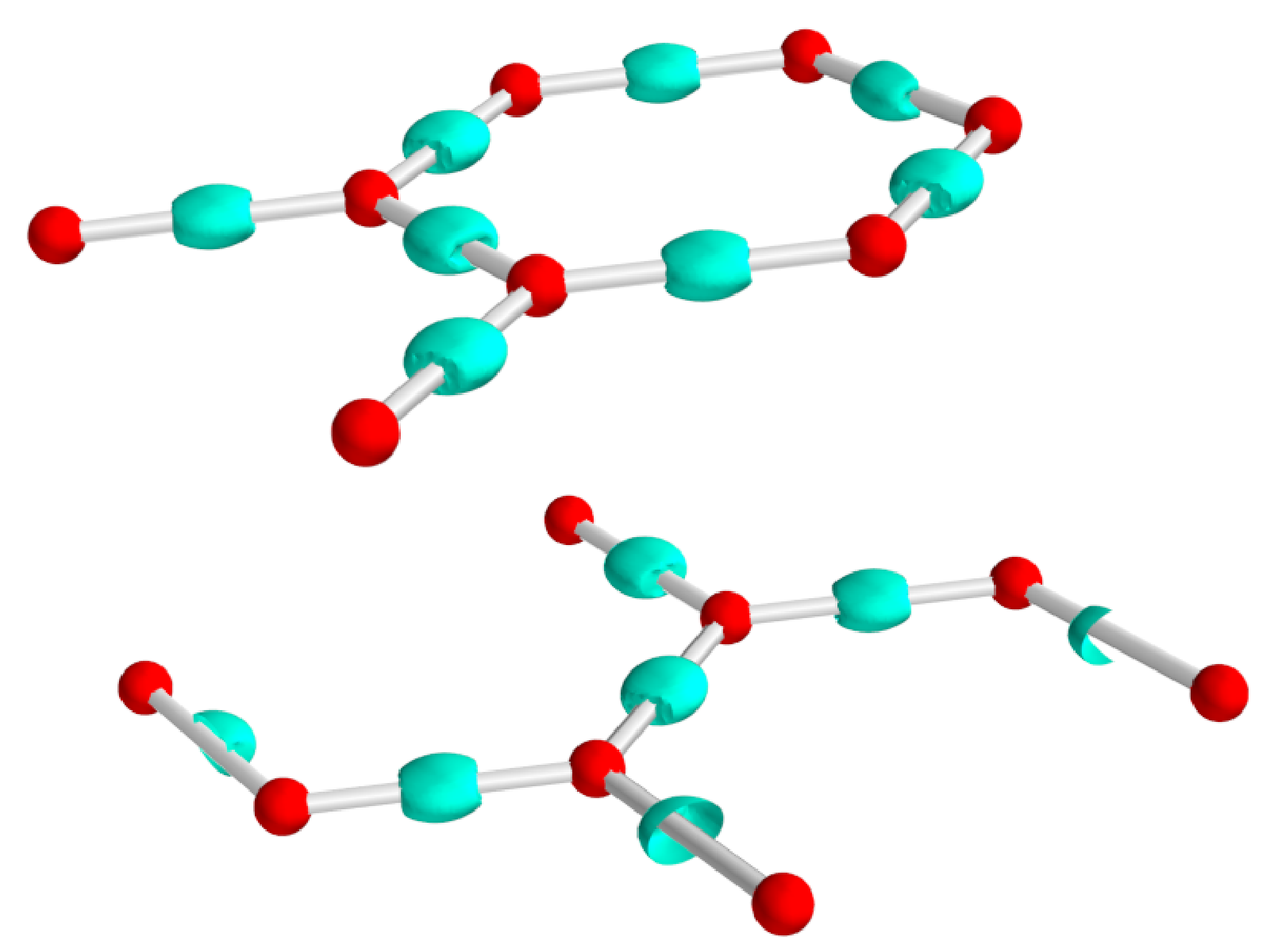}}
\put(0,0){\epsfxsize=7cm \epsfbox{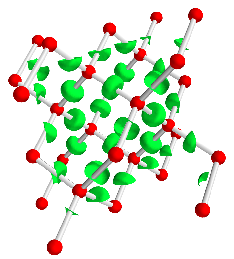}}
\put(0,13){(a)}
\put(0,6){(b)}
\end{picture}
\caption{Real space position of the lumps of Fig.~\ref{fig:all}.
The atoms are represented by red spheres, while the erroneous points for
(a) graphite (isosurface corresponding to
$t^{\text{GEA2L}}/t^{\text{KS}}=2.25$) and (b) silicon
(isosurface corresponding to $t^{\text{GEA2L}}/t^{\text{KS}}=1.9$) are in
turquoise and green, respectively.}
\label{fig:3ds}
\end{figure}

In Fig.~\ref{fig:3ds}, the erroneous points from these two regions are shown in
real space, where we can see that the bigger errors occur in the middle of covalent
bonds.
If, for instance, for graphite the same PCA method is applied using only the points
in the bonding regions, a much better accuracy (in these bonding regions) can be reached,
and the resulting linear combination is given by
\begin{equation}
t^{\text{PCA}}_{\text{bond}}(\mathbf{r}) = 
0.389t^{\text{TF}}(\mathbf{r}) +
0.635t^{\text{W}}(\mathbf{r})+0.084\nabla^2\rho(\mathbf{r}).
\end{equation}
While this is obviously not useful as a general KED approximation,
it is interesting to note that $t^{\text{W}}$ has now a small positive coefficient,
in agreement with the fact that the covalent $\sigma$-bonding
in graphite and silicon should be dominated by a single molecular orbital.
As shown by Seino \textit{et al}.,\cite{SeinoJCP18} considering also
the third derivative of $\rho$ further improves the accuracy of OF KED.
However, as discussed below, the bonding regions highlighted in Fig.~\ref{fig:3ds}
are not necessarily those which are the most relevant for explaining the differences
observed in the results for the lattice constant.

\begin{figure}
\includegraphics[width=1.0\columnwidth]{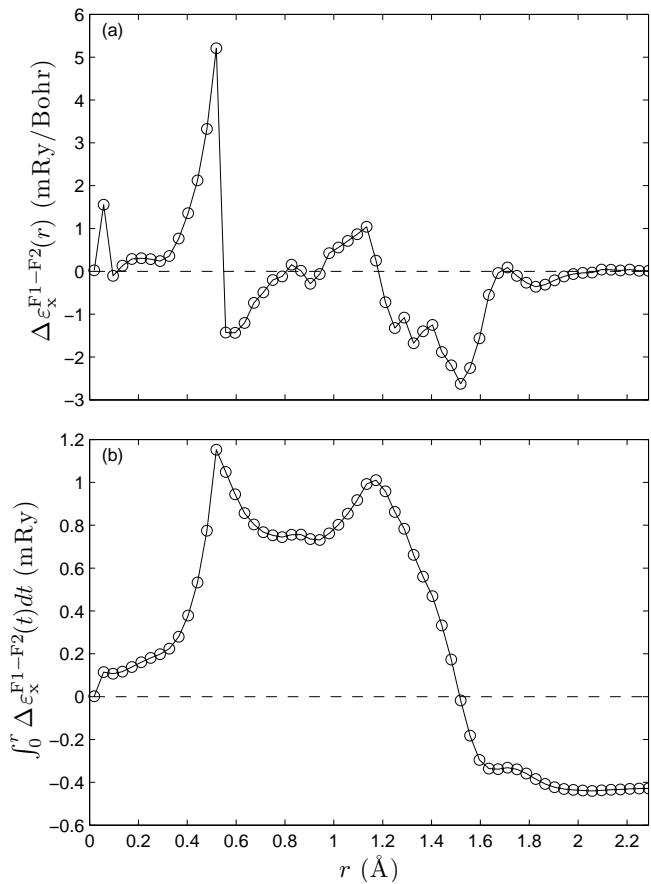}
\caption{Difference between the exchange components of SCAN(CRopt) (F1)
and SCAN(GEA2L) (F2) in Si plotted as a function of the distance $r$ from
an Si atom. Panel (a) shows the angular average of
$\Delta\varepsilon_{\text{x}}^{\text{F1-F2}}$ (see text for definition), while
panel (b) shows the radial integration of $\Delta\varepsilon_{\text{x}}^{\text{F1-F2}}$
from the atom until $r$.}
\label{fig_CRopt_GEA2L}
\end{figure}
\begin{figure}
\includegraphics[width=1.0\columnwidth]{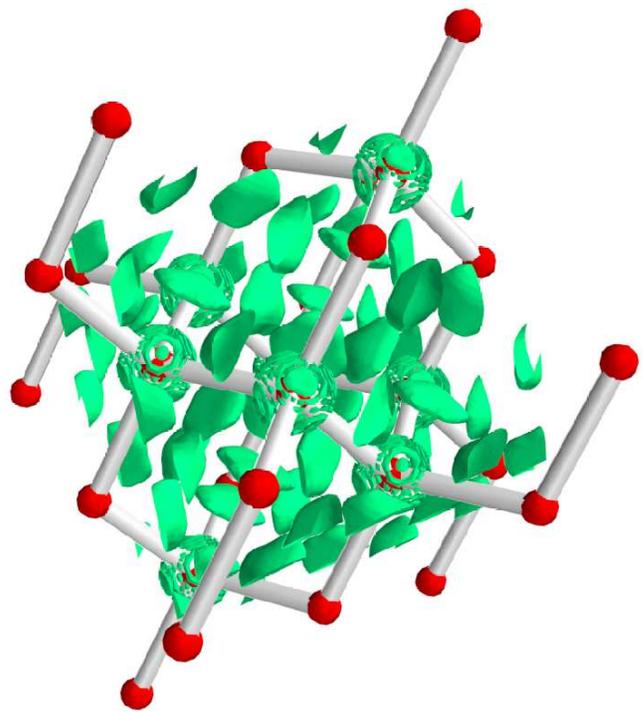}
\caption{Isosurface of the absolute value of
$F_{\text{x}}^{\text{SCAN(CRopt)}}-F_{\text{x}}^{\text{SCAN(GEA2L)}}$ corresponding to 0.03.}
\label{Si_enhancement_CR_GEA_0.03}
\end{figure}
In order to provide some insight into the results presented in
Sec.~\ref{results}, Fig.~\ref{fig_CRopt_GEA2L} compares the energy density
of SCAN(GEA2L) and SCAN(CRopt) in Si. For simplicity, only the exchange
component, which is much larger than correlation, is considered.
SCAN(GEA2L) and SCAN(CRopt) lead to rather different equilibrium lattice
constants $a_{0}$ for Si, namely, 5.437 and 5.460~\AA, respectively,
and the following analysis provides details about the regions of space that are
involved to explain these different values of $a_{0}$.
Figure~\ref{fig_CRopt_GEA2L}(a) shows
$\Delta\varepsilon_{\text{x}}^{\text{F1-F2}}$, which is defined as
\begin{eqnarray}
\Delta\varepsilon_{\text{x}}^{\text{F1-F2}}(r) & = &
r^{2}\int
\left[
\left(\varepsilon_{\text{x}}^{\text{F1},a_{\text{large}}}(\mathbf{r}) -
\varepsilon_{\text{x}}^{\text{F1},a_{\text{small}}}(\mathbf{r})\right)\right. \nonumber \\
& & - \left.\left(\varepsilon_{\text{x}}^{\text{F2},a_{\text{large}}}(\mathbf{r}) -
\varepsilon_{\text{x}}^{\text{F2},a_{\text{small}}}(\mathbf{r})\right)\right]d\Omega,
\label{Deltaexc}
\end{eqnarray}
where $\varepsilon_{\text{x}}^{\text{F},a}$ is the exchange energy density
[defined by Eq.~(\ref{ExcMGGA})]
of functional F (F1 and F2 designate SCAN(CRopt) and SCAN(GEA2L), respectively)
calculated at a given lattice constant
($a_{\text{small}}$ or $a_{\text{large}}$). The integration in Eq.~(\ref{Deltaexc})
is over the spherical angles and $r$ is the distance from one Si atom.
As discussed in detail in Refs.~\onlinecite{HaasPRB09b,LevamakiPRB14}, the
equilibrium lattice constant $a_{0}$
is determined by the slope of the xc-energy $E_{\text{xc}}$, i.e.,
the variation of $E_{\text{xc}}$ with respect to $a$, and this is basically
what Fig.~\ref{fig_CRopt_GEA2L} shows since the difference between two values of
$a$ ($a_{\text{small}}$ and $a_{\text{large}}$) is considered. Figure~\ref{fig_CRopt_GEA2L}(b) shows
the radial integration of $\Delta\varepsilon_{\text{x}}^{\text{F1-F2}}$ up to
a given value of $r$. As already discussed in Ref.~\onlinecite{HaasPRB09b}
for Si but in the case of GGA functionals,
two different regions contribute significantly to the variation of $E_{\text{xc}}$
with respect to $a$. The first one, located around
0.5~\AA~[see the fast variations of the curves in Figs.~\ref{fig_CRopt_GEA2L}(a) and (b)]
corresponds to the core-valence separation. The second region extends from
1.2 to 1.7~\AA~and corresponds to the valence/interstitial region which is rather
large since Si has an open structure.
Additionally, Fig.~\ref{Si_enhancement_CR_GEA_0.03} shows the isosurface of
$\left\vert F_{\text{x}}^{\text{SCAN(CRopt)}}-F_{\text{x}}^{\text{SCAN(GEA2L)}}\right\vert$
that delimits values larger than 0.03
(where actually $F_{\text{x}}^{\text{SCAN(CRopt)}}>F_{\text{x}}^{\text{SCAN(GEA2L)}}$)
and highlights the two types of regions just mentioned above.

\begin{figure}
\includegraphics[width=1.0\columnwidth]{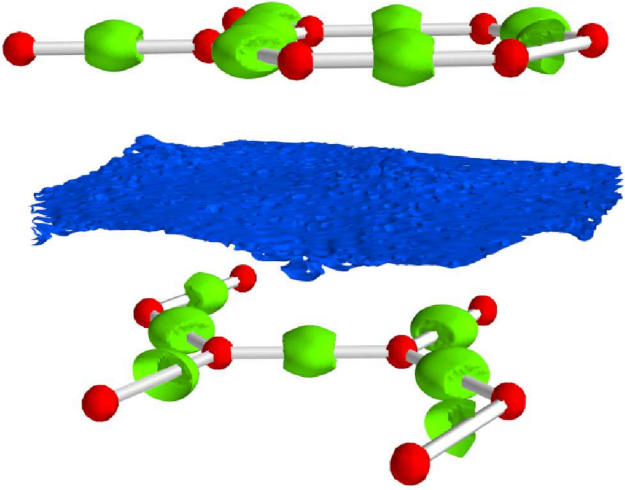}
\caption{The regions of space in graphite where
$t^{\text{GEA2L}}/t^{\text{KS}}$ and $t^{\text{KS}}/t^{\text{GEA2L}}$
are larger than 1.9 are delimited by the isosurfaces in green and blue,
respectively.}
\label{Graphite_GEA_1.9_colored}
\end{figure}
The case of graphite was also discussed in Ref.~\onlinecite{HaasPRB09b},
where the electron density and reduced density gradient $s$ in the
region between the layers were studied in detail. It was shown
that an increase of the interlayer distance leads to a rather large
increase of $s$ overall, thus explaining the overestimation of the
interlayer distance for GGA functionals with a too strong enhancement factor.
Figure~\ref{Graphite_GEA_1.9_colored} shows the ratio $t^{\text{GEA2L}}/t^{\text{KS}}$
with a ratio that is smaller than the one used in Fig.~\ref{fig:3ds}(a), such
that the isosurface encloses a larger region. We can see that aside from the
middle of the short covalent bonds within the layers (not relevant for
the interlayer distance), also a non-negligible portion of the
space between the layers has a ratio
($t^{\text{KS}}/t^{\text{GEA2L}}$) bigger than 1.9.

\begin{figure}
\includegraphics[width=1.0\columnwidth]{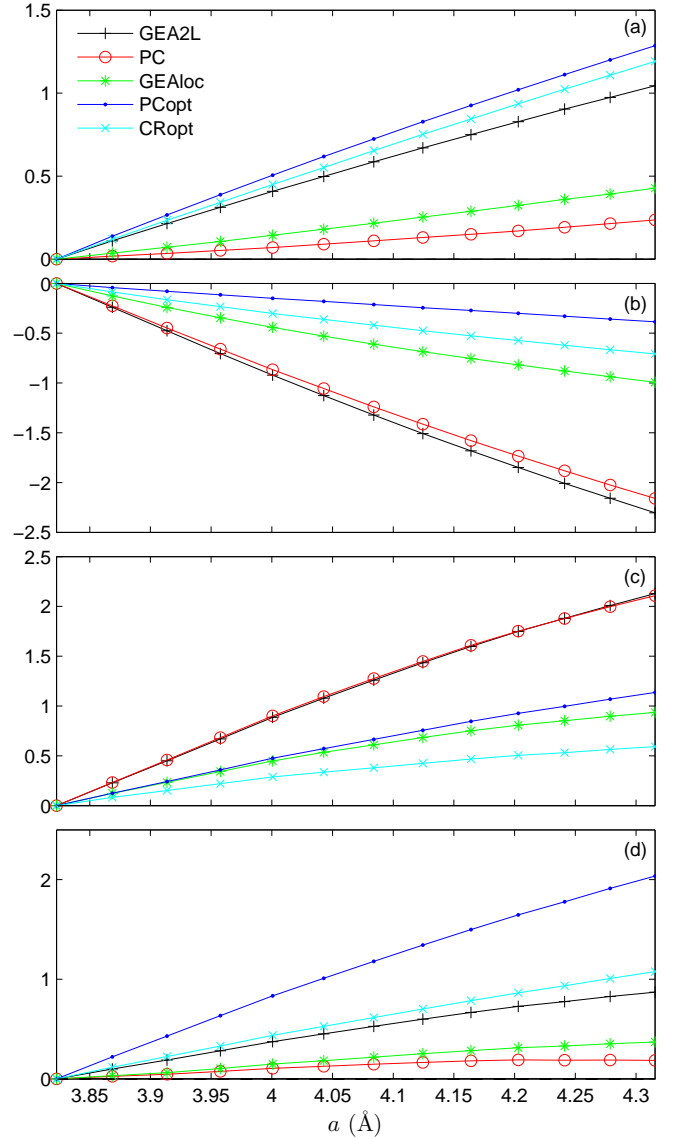}
\caption{Difference
$E_{\text{xc}}^{t^{\text{OF}}\text{-SCAN}}-E_{\text{xc}}^{t^{\text{KS}}\text{-SCAN}}$
(in mRy) between the xc-energies of LiH obtained
with SCAN and its deorbitalized versions plotted as a function of the
lattice constant $a$. Panels (a), (b), and (c) show the contributions from the
Li atom, H atom, and interstitial region, respectively, while panel (d)
shows the sum of all contributions (i.e., the whole unit cell).
The atomic muffin-tin spheres of the Li and H atoms are 1.7~Bohr.
Each curve is vertically shifted such that the zero is at the smallest volume.}
\label{fig_EXC_LiH_SCAN}
\end{figure}

\begin{figure}
\includegraphics[width=1.0\columnwidth]{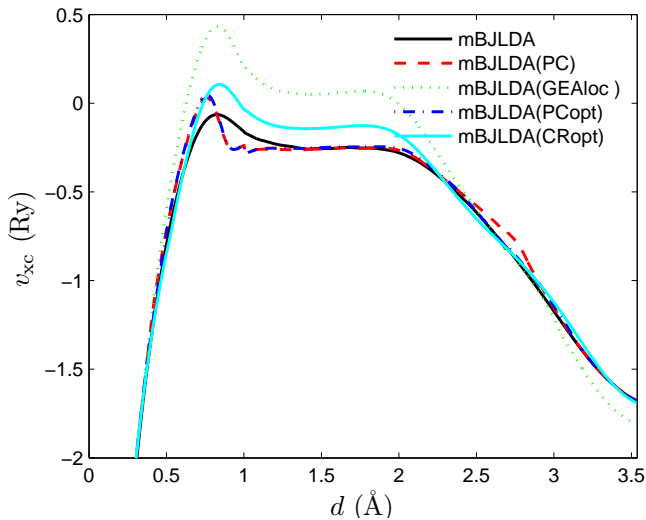}
\caption{mBJLDA xc-potential and a few selected of
its deorbitalized versions plotted in LiH from the Li atom at
(0,0,0) to the H atom at $(\frac{1}{2},\frac{1}{2},\frac{1}{2})$.}
\label{fig_vxc_LiH}
\end{figure}

In Sec.~\ref{results}, we also observed
that an OF KED that is among the most accurate for a property calculated
with the total energy, may be among the most inaccurate for the band gap,
or vice versa. For instance, while $t^{\text{PCopt}}$ and $t^{\text{CRopt}}$
are not among the best KEDs for total-energy related properties of strongly bound
solids, they are the most accurate for the band gap. Such contradictory results
could seem quite puzzling at first sight, however this should be rather simple
to explain in most cases.

Taking LiH as an example, Fig.~\ref{fig_EXC_LiH_SCAN} compares the xc-energy
calculated with SCAN and selected deorbitalized SCANs by showing the difference
$E_{\text{xc}}^{t^{\text{OF}}\text{-SCAN}}-E_{\text{xc}}^{t^{\text{KS}}\text{-SCAN}}$
as a function of the lattice constant $a$ (this is the same kind of analysis
as the one used for Si in Fig.~\ref{fig_CRopt_GEA2L}).
Figures~\ref{fig_EXC_LiH_SCAN}(a), \ref{fig_EXC_LiH_SCAN}(b), and
\ref{fig_EXC_LiH_SCAN}(c) show the contributions from the
Li atom, H atom, and interstitial region, respectively, while the
sum of them (the total value in the unit cell)
is shown in Fig.~\ref{fig_EXC_LiH_SCAN}(d). As expected, the SCAN
equilibrium lattice constants $a_{0}$ of LiH (see Table~S2 of the supplementary
material\cite{SM_tau}) show the same ordering as the curves in
Fig.~\ref{fig_EXC_LiH_SCAN}(d) [the uppermost (lowest) curve correspond
to the smallest (largest) lattice constant].
Thus, in the present case where the same functional is evaluated with different
KED, the change in $a_{0}$ due to deorbitalization
depends on the variation with $a$
of the difference between $t^{\text{KS}}$ and $t^{\text{OF}}$.
From Fig.~\ref{fig_EXC_LiH_SCAN}, we can also see that for all functionals,
$E_{\text{xc}}^{t^{\text{OF}}\text{-SCAN}}-E_{\text{xc}}^{t^{\text{KS}}\text{-SCAN}}$
decreases in the H atom, but increases in the Li atom and interstitial
region such that in total an increase is obtained.
We also note that with $t^{\text{GEA2L}}$ and $t^{\text{PC}}$ there is a very
large cancellation of the errors coming from the H atom and interstitial region.

As discussed in previous works,\cite{StaedelePRL97,TranJPCM07,TranJCTC15}
the magnitude of the band gap is determined by the inhomogeneities in the potential,
such that, roughly speaking, large inhomogeneities favor larger values of the band gap.
Actually, in most systems the valence band maximum and conduction band
minimum are located close to an atom and in the interstitial region, respectively,
which means that the difference in the magnitudes of a potential
between these two regions determines the band gap.
Again for LiH, Fig.~\ref{fig_vxc_LiH} compares $v_{\text{xc}}$ of
mBJLDA and its OF variants. The LiH band gap (see Table~S11 of the supplementary
material\cite{SM_tau}) with mBJLDA is 5.06~eV and is reproduced at best
by mBJLDA(PCopt) (5.03~eV), while mBJLDA(GEAloc) with 6.69~eV leads to the
worst agreement. This is in accordance with Fig.~\ref{fig_vxc_LiH}, where
we can see that the mBJLDA(PCopt) potential is the closest to
mBJLDA, while the mBJLDA(GEAloc) potential is much higher
in the interstitial region (where the conduction band minimum is located)
and lower close to the H atom (where the valence band maximum is located).

Thus, from this detailed discussion about LiH it is rather clear that
different mechanisms have to be invoked in order to explain the trends observed
for the lattice constant (a total-energy related property) and band gap, such
that opposite conclusions for these two types of properties can be obtained.

\section{\label{summary}Summary}

In this work, the deorbitalization of several xc-MGGA methods, three energy
functionals and one potential, has been investigated by considering properties
of solids. The replacement $t^{\text{KS}}\rightarrow t^{\text{OF}}$
in xc-MGGAs affects the results to some degree which depends on both the
xc-MGGA under investigation and the used approximation for the OF KED
$t^{\text{OF}}$.

Concerning the energy functionals for the calculation of the lattice constant,
bulk modulus, and binding energy, we have shown that the results
are in general more sensitive with MVS than with SCAN and TM, which
should just be the direct consequence of the analytical form of the functionals
that depends more strongly on the KED in the case of MVS.
With SCAN and TM, the replacement
$t^{\text{KS}}\rightarrow t^{\text{OF}}$ with most OF KED
does not change much the results for strongly bound solids, such that the
performance of a xc-MGGA remains pretty much the same. For the weakly bound
rare gases, the change in the cohesive energy is usually rather
large, while for the layered solids large changes in the interlayer
distance are obtained with MVS.

The deorbitalization of the mBJLDA xc-potential leads to appreciable
changes in the band gap and only the OF KED $t^{\text{PCopt}}$ can be
considered as a somehow reasonable replacement of $t^{\text{KS}}$.

Similarly as Mejia-Rodriguez and Trickey,\cite{MejiaRodriguezPRA17} we were
not able to identify a OF KED that leads to reasonably small change in the
results in most circumstances.

\begin{acknowledgments}

This work was supported by the project F41 (SFB ViCoM) of the Austrian Science
Fund (FWF) and by the TU-D doctoral college (TU Wien).
F.T. acknowledges discussions with J. P. Perdew and L. A. Constantin.

\end{acknowledgments}

\bibliography{references}

\end{document}